# Sub-optical wavelength acoustic wave modulation of integrated photonic resonators at microwave frequencies

Semere Ayalew Tadesse[1,2] and Mo Li[1*]

[1] *Department of Electrical and Computer Engineering, University of Minnesota, Minneapolis, MN 55455, USA*

[2] *School of Physics and Astronomy, University of Minnesota, Minneapolis, MN 55455, USA*

**Light-sound interactions have long been exploited in various acousto-optic devices based on bulk crystalline materials. Conventionally these devices operate in megahertz frequency range where the acoustic wavelength is much longer than the optical wavelength and a long interaction length is required to attain significant coupling. With nanoscale transducers, acoustic waves with sub-optical wavelengths can now be excited to induce strong acousto-optic coupling in nanophotonic devices. Here we demonstrate microwave frequency surface acoustic wave transducers co-integrated with nanophotonic resonators on piezoelectric aluminum nitride substrates. Acousto-optic modulation of the resonance modes at above 10 GHz with the acoustic wavelength significantly below the optical wavelength is achieved. The phase and modal matching conditions in this scheme are investigated for efficient modulation. The new acousto-optic platform can lead to novel optical devices based on nonlinear Brillouin processes and provides a direct, wideband link between optical and microwave photons for microwave photonics and quantum optomechanics.**

## Introduction

Actively generated acoustic wave in optical materials, typically quartz, lithium niobate and tellurium dioxide, can act as a traveling phase grating to deflect incident lightwave through Bragg diffraction and shift its frequency through the Doppler effect[1]. These mechanisms have been applied to build an important family of optical devices—acousto-optic (A/O) devices—that include modulators, frequency shifters, beam deflectors and scanners, tunable filters, spectral analyzers and Q-switches in pulsed lasers[2]. In addition to diffraction effects, nonlinear optical effects such as stimulated Brillouin scattering (SBS) and amplification, a three-wave mixing process between light and sound waves, are also in the realm of acousto-optics[3-6]. Conventional A/O devices based on bulk crystalline materials, however, are bulky and their integration with highly integrated optical systems such as silicon photonics remains to be a challenge. Moreover, their operation frequencies are limited to megahertz range and thus insufficient for modern optical communication operating with bandwidth of many gigabits/sec. Past efforts to increase the frequency of A/O devices are hindered by the fabrication capability of acoustic transducers and the much reduced elasto-optic efficiency at higher frequencies.

* Corresponding author: moli@umn.edu

To integrate acousto-optics, a new paradigm was introduced in the seventies which integrated surface acoustic wave (SAW) transducers with planar optical waveguides[7,8]. In contrast to bulk acoustic waves, surface acoustic waves propagate on the surface of piezoelectric materials in acoustic modes confined within a depth on the order of the acoustic wavelength[9,10]. SAW thus can have very high energy confinement and overlap with the optical modes of the planar waveguides to achieve efficient A/O modulation in a compact device. However, the highest operation frequency of such SAW based integrated A/O devices that have been achieved was still below one GHz so far[11-15].

With the significant advances in nanofabrication, inter-digital acoustic transducers can now be readily fabricated with sub-micron linewidth to generate surface acoustic waves with ultrahigh frequency up to tens of GHz[16-22]. At the same time, nanophotonic waveguides and cavities with very high quality factors have been developed to confine light in sub-wavelength scale with extremely high optical power density[23,24]. Combining above two advances, acousto-optics can enter an unprecedented regime in which the acoustic wavelength ($\Lambda = c_a/f_a$; $c_a$: sound velocity; $f_a$: acoustic frequency) can be reduced to much less than the optical wavelength $\lambda/n$ ($n$: refractive index). In this sub-optical wavelength regime, nearly ideal mode overlapping and phase matching conditions between light and sound waves can be reached in a highly confined system to attain efficient acousto-optic interaction and induce strong nonlinear effects such as Brillouin scattering. Indeed, gigahertz surface acoustic wave has recently been used to modulate the emission of GaAs quantum dots embedded in a photonic crystal nanocavity with the modulation frequency reaching a record of 1.7 GHz[25]. The acoustic wavelength (~1.8 μm) generated in that device, however, is still many times of the optical wavelength (~0.25 μm in GaAs). Here we demonstrate acoustic modulation of photonic cavities with surface acoustic wave of frequency up to 10.6 GHz. In this microwave X-band frequency range, the acoustic wavelength is reduced to less than the optical wavelength (~0.75 μm), for the first time to the best of our knowledge, reaching the sub-optical wavelength regime of integrated acousto-optics.

**Results**

**Acousto-optic overlap in sub-optical wavelength regime**

The proposed devices integrate both surface acoustic wave and photonic devices on 330 nm polycrystalline aluminum nitride (AlN) films sputtered on silicon wafers with 3 μm thick thermal oxide layers. Given its strong piezoelectricity, high sound velocity and high refractive index, AlN is an ideal active material for both the excitation of acoustic waves and the making of optical waveguides[26-28]. In the highly integrated acousto-optic devices, the wavelength Λ of ultrahigh frequency acoustic waves can be substantially smaller than the size of the optical mode in the waveguides. In this regime, the spatial overlap between the acoustic wave and the optical mode is critical to the efficiency of acousto-optic modulation and is more sophisticated than that in conventional A/O devices. Fig. 1a schematically illustrates the relative scales of the acoustic wave and the optical mode in such a regime. The acoustic wave generates a propagating strain field, tensile and compressive periodically, which modulates the phase of the optical mode by changing the refractive index through a combination of elasto-optic and electro-optical effects. As a result, the resonance frequency of the resonator is modulated which can be approximated by using the perturbation theory as[29]:

$$\frac{\Delta\omega}{\omega} \approx -\frac{\iint \Delta n(\mathbf{r})\left|E(\mathbf{r})\right|^2 d\mathbf{r}}{\iint n(\mathbf{r})\left|E(\mathbf{r})\right|^2 d\mathbf{r}} \qquad (1)$$

where $\Delta n$ is the elasto-optic change of the refractive index tensor $\Delta n_i = -\frac{1}{2}n_i^3\sum_j p_{ij}S_j \ \left(i, j = 1, 2 \ldots 6\right)$, with $p_{ij}$ and $S_j$ representing the contracted effective elasto-optic coefficient tensor and the strain field tensor, respectively[30]. Taking into account the polycrystalline nature of the AlN film and its contracted elasto-optic coefficient tensor, equation

(1) can be reduced to a scalar expression which is proportional to an overlap factor $\Gamma$ representing the spatial overlap between the strain field of the acoustic wave and the electrical field of the waveguide mode (see Supplementary Note 2 and 3):

$$\Gamma = \frac{\iint \left[ p'_{11} S_1(x,z) + p_{13} S_3(x,z) \right] |E_x(x,z)|^2 \, dxdz}{\iint |E_x(x,z)|^2 \, dxdz} \tag{2}$$

This overlap factor $\Gamma$ critically determines the acousto-optic modulation efficiency. Simulation shows that $\Gamma$ strongly depends on the acoustic wavelength as it is reduced to be less than the waveguide width $W$. Fig. 1b illustrates the simulated field distribution of the optical mode, the acoustic mode and their spatial overlap, which is integrated as in equation (2) to give $\underline{\Gamma}$, in two representative situations when $\Lambda=2W$ and $\Lambda=W/2$, respectively. The simulation predicts that optimal overlap is achieved when the acoustic wavelength $\Lambda$ is close to twice of the lateral size of the optical mode. In contrast, when the acoustic wavelength almost equals the mode size, the modulation by tensile and compressive strain will nearly cancel each other, leading to vanishing modulation. The situation varies periodically as $\Lambda$ is continuously reduced for a given waveguide size. In the following we experimentally verify these relations in devices with varying acoustic wavelength $\Lambda$ and optical waveguide width $W$.

**Device Details**

Fig. 1c and d shows optical and scanning electron microscopy images of a typical device, which features an inter-digital transducer (IDT) made of gold and an optical racetrack resonator made of AlN single-mode rib waveguide. Among the devices studied, the width of the waveguide is varied from 0.8 to 1.2 μm whereas the etched depth is constantly 200 nm. The IDTs have period $D$ in the range of 0.4 to 4 μm, which corresponds to the wavelength ($\Lambda$) of the acoustic wave they excite. Fig. 1d shows a zoom-in image of an IDT with $D$=0.4 μm; the width of each electrode finger is $D/4$=100 nm. The IDTs are placed to launch acoustic waves propagating in the direction transverse to the straight waveguide segments of the racetrack (Fig. 1c), which are designed to have the same length as the aperture ($A$) of the IDTs in order to achieve maximal acousto-optic interaction. The optical resonances of the racetrack resonator can be observed in the transmission spectrum measured from the feeding waveguide as shown in Fig. 1e. The best intrinsic optical quality factor with waveguide width of 0.8 μm is $8\times10^4$.

**SAW Characteristics**

We first characterized the SAW devices by conducting microwave reflection measurements. The SAW IDT was contacted with a microwave coplanar probe and the reflection coefficient $S_{11}$ of the device was measured with a network analyzer. In Fig. 2a, we show a series of $S_{11}$ parameters versus frequency measured from IDTs with $\Lambda$=1.6, 0.9, 0.7 and 0.5 μm. Acoustic modes with frequency up to 12 GHz into the microwave X-band can be observed in the reflection spectra as prominent dips. Numerical simulation confirms that these are the Rayleigh modes of different orders with the mode number $n$ up to 12. They are labeled as $R$n in Fig. 2a. It can be noted that for a given $\Lambda$, only a selective set of modes are excited and as $\Lambda$ reduces, low order modes start to disappear and higher order modes begin to dominate. Simulation confirms that these are the allowable modes restricted by the boundary conditions of the AlN/SiO₂/Si multilayers. The multilayer structure and the relatively thin AlN layer, as is necessary for making the photonic devices, also leads to strong dispersion of the acoustic waves because the modes extend over the layers with different sound velocities. Therefore, the dispersion relation between the phase velocities and the wavenumber of the acoustic waves is very important to the design of these acoustic devices. We experimentally determined the phase velocity of different modes from their frequency as $u_\mathrm{p} = \omega_\mathrm{a}/k_\mathrm{a}$, where $\omega_\mathrm{a}$ is the angular frequency and $k_\mathrm{a} = 2\pi/\Lambda$ is the wavenumber of the acoustic modes. The results are displayed in Fig. 2b along with the dispersion curves calculated with numerical simulation. The experimental and the theoretical results show reasonable agreement with discrepancies attributed to the finite simulation space and the mass loading effect of the IDT electrodes. Because of their different mode profiles (see Supporting Information), different orders of Rayleigh modes show very distinct dispersion properties. As can be seen in Fig. 2b, with reducing wavenumber the phase velocities of the low order modes approach the sound velocity of SiO₂ ($c_{\mathrm{SiO_2}}$ = 3400 m/s) because the acoustic modes largely reside in the SiO₂ layer. In contrast, the phase velocities of the high order modes approach the sound velocity of AlN ($c_{\mathrm{AlN}}$ =6000 m/s) because the acoustic mode is more confined in the AlN layer. Another important parameter of acoustic transducers is the electromechanical coupling coefficient $k^2$, which is defined as the ratio of the mechanical power of the acoustic wave and the input electrical power, and thus characterizes the transducers' energy efficiency. In Fig. 2c, $k^2$ of the fundamental Rayleigh mode ($R$1) is plotted versus wavelength $\Lambda$, showing that at wavelength above 1 μm $k^2$ is around 0.4%, which is comparable with values of SAW devices made in similar thin films[31,32]. However, when Λ is reduced to below 1 μm, $k^2$ decreases dramatically, making the fundamental mode more difficult to be excited as observed in Fig. 2a. On the other hand, in Fig. 2d the measured $k^2$ of different mode orders are compared for two fixed wavelengths (Λ=1.6 and 0.85 μm). The result reveals that at very short wavelength $k^2$ recovers for higher mode orders and becomes comparable with that of the fundamental mode at long wavelengths. Above results suggest that the high order modes are the dominant modes at the limit of short acoustic wavelengths (where $k_a h > 1$, $h$ is the AlN film thickness) and advantageous in excitation efficiency and reaching ultrahigh frequency. In the following, we demonstrate ultrahigh frequency acousto-optic modulation of photonic resonators using these high order acoustic modes.

**Surface acousto-optic modulation**

To measure the acousto-optic modulation, a tunable laser is coupled into the feeding waveguide and the racetrack resonator through integrated grating couplers and the transmission out of the device is monitored with a high-speed photodetector connected to the port 2 of the network analyzer. With this configuration, the measured spectra of the transmission coefficient $S_{21}$ reflect the frequency response of acousto-optic modulation in the system. To convert the acousto-optic phase modulation to amplitude response of the transmitted optical signal, the input laser is detuned from the optical resonance to implement the slope detection scheme. In Fig. 3a the results obtained from the same set of devices as in Fig. 2a are displayed. The width of the optical waveguide $W$ is fixed at 0.8 μm in these devices. In the spectra of $S_{21}$, peaks induced by the acoustic modes with frequencies that match those in the $S_{11}$ spectra of Fig. 2a can be observed. The highest modulation frequency reaches 10.6 GHz for the 7th Rayleigh mode with acoustic wavelength Λ=0.5 μm. This demonstrated acousto-optic modulation frequency is almost an order of magnitude higher than the previously reported result[23]. More importantly, our results represent the first demonstration of acousto-optic modulation in integrated photonic devices using acoustic waves with wavelength significantly smaller that the optical wavelength (~0.75 μm here). The modulation amplitude was also measured with varying laser detuning relative to the optical resonance frequency. The result for the R6 mode of Λ=0.9 μm device is displayed in Fig. 3b, showing that the optical $S_{21}$ follows the derivative of a Lorentzian optical resonance lineshape, as expected from the phase modulation nature of the acousto-optic interaction. In addition, because the phase shift induced by the acousto-optic modulation is expected to be proportional to the amplitude of the acoustic wave, the modulation amplitude should be linearly dependent on the square root of the input electrical power. This linear relation is clearly shown in Fig. 3c, measured from three representative acoustic modes (R3, R4, R5) of the $\Lambda$=0.9 μm device. The actual input power $P$ is calculated by taking into account the reflection coefficient of the IDT and the modulated optical power is calculated from the $S_{21}$ coefficient after calibrating the gain and losses of all the optical and electrical elements in the setup (see Supplementary Note 5).

Comparing Fig. 2a and Fig. 3a, it can be observed that the electrical reflection and the optical transmission measurements yield different relative amplitudes for different acoustic modes. For example, the R11 mode of the $\Lambda$=0.5 μm device is strong in the electrical $S_{11}$ spectrum but suppressed in the optical $S_{21}$ spectrum. It indicates that even though this mode can be efficiently excited, it does not modulate the optical mode effectively as compared with, for example, the R9 mode of the same device. This effect is attributed to the different acousto-optic overlap factors $\Gamma$ for different modes, as well as their different electromechanical coupling efficiencies $k^2$. $\Gamma$ is strongly dependent on both the acoustic wavelength $\Lambda$ and the perpendicular mode profile of the acoustic mode (Fig. 1b) for a given optical waveguide size. As $\Lambda$ is reduced to below the optical wavelength, the strain field across the optical mode can no longer be approximated as uniform but rather is spatially periodic as illustrated in Fig. 1. Therefore to evaluate the optomechanical modulation of the optical resonance, $\Gamma$ has to be calculated for each given acoustic wavelength and optical waveguide size. To be consistent with other types of optomechanical systems, we define the opto-mechanical coupling coefficient $G_{om} = \Delta\omega/\Delta z$ as the ratio of the optical resonance frequency shift to the displacement amplitude ($\Delta z$) of the acoustic wave. $G_{om}$ describes the efficiency of acousto-optic modulation in a specific device and can be determined experimentally (see Supplementary Note 3). The representative results of $G_{om}$ for the second (R2) and the sixth (R6) Rayleigh modes measured from eighteen devices with different $\Lambda$ and waveguide widths ($W$=0.8μm, 1.05μm and 1.2μm) are shown as symbols in Fig. 4a. For comparison, the calculated overlap factors divided by $k^2$ to compensate for the different electromechanical coupling efficiencies are also plotted after being normalized. Qualitative agreement between the theory and the experiment can be observed. The results show that $G_{om}$ reaches maximal value when $W$ is smaller than the acoustic wavelength $\Lambda$ and the effective width of the optical mode is close to $\Lambda/2$.

In contrast, as $W$ increases above $\Lambda$, $G_{om}$ first decreases and then oscillates when the overlap integral changes periodically. Thus, to achieve efficient acousto-optical modulation, the optical waveguide width needs to be optimally designed to utilize a specific acoustic mode. In addition to the lateral modal overlap, the acousto-optic overlap in the vertical direction, which depends strongly on the mode orders, also plays an important role. In Fig. 4c, we show the measured $G_{om}$ of different acoustic modes in a $\Lambda$=0.9μm, $W$=0.8 μm device along with theoretical results. The experimental results agrees well with the theoretical prediction to show that at this particular wavelength, the fundamental and the third order modes are the most efficient in acousto-optical modulation but the third order mode can modulate at higher frequency with similar efficiency. As shown in Fig. 2d, when the acoustic wavelength is further reduced, the higher order modes can be more efficient and thus more desirable for ultrahigh frequency modulation.

**Wide band acousto-optic modulation**

An important application of SAW devices is for front-end bandpass filters in wireless communication systems which often requires a wide passband width[9]. Wideband SAW transducer can be achieved with a slanted IDT design in which the period of IDT electrodes is varied linearly to generate SAW wave with frequencies spanning between the frequencies that correspond to the periods at the two ends of the IDT[33,34]. To demonstrate broadband acousto-optic modulation, we designed a slanted IDT with a center period of 4μm (corresponding center frequency: 0.9 GHz) and 20% bandwidth. The acousto-optic modulation response of the device is shown in Fig. 5a. The result demonstrates a −3dB bandwidth of 130 MHz and a side-lobe rejection ratio of over 30dB. In the passband, however, strong passband ripples can be observed. These ripples are attributed to the phase delay between the modulation at the back and front straight segments of the racetrack due to the propagation of acoustic wave. As illustrated in Fig. 5a, the peak modulation amplitude occurs when the phase delay $\Delta\phi = 2\pi f \Delta L/u_p$, where $\Delta L$=300 μm is the distance between the waveguides, $u_p$ is the phase velocity, equals $2N\pi$ ($N$, an integer) and thus the modulations at the two waveguide segments are in phase. Therefore, the frequencies at the peaks of the passband ripples are related to the phase velocity of the acoustic wave as $f_N = N u_p/\Delta L$. Linear fitting of the measurement result as shown in Fig. 5b yields a phase velocity value of 3400 km/s, which agrees well with the result for the R1 mode of the same acoustic wavelength (Fig. 2b). In addition to phase velocity, the broadband result can also reveal the propagation loss of the acoustic wave.

Similar to other types of delay line interferometers, the extinction ratio $\eta$, defined as the ratio of the maximum and the minimum of the ripples, is related to the amplitude propagation loss $\alpha$ by $\exp(-\alpha \Delta L) = (\eta - 1)/(\eta + 1)$. The result of $\alpha$ determined with this method is shown in Fig. 5c. An average value of 6 dB/mm is obtained, which should be considered as an upper bound estimation of the propagation loss because effects such as the reflection of the acoustic wave by the waveguide are not considered in the calculation. With this apparently high value of linear loss[35,36], the total acoustic loss, however, is insignificant considering that the footprint of photonic devices typically spans only a few tens of micrometers. Although the interference effect can be utilized to enhance acousto-optic modulation at a specific frequency, the large passband ripples are undesirable for signal processing applications. They can be easily removed by including an acoustic absorber in the center of the racetrack resonator to allow only the front waveguide segment to be modulated. Further optimization of the slanted IDT design can be done to improve its passband width, shape factor and stopband suppression so the device may be applied as an efficient broadband filter in radio-frequency photonics systems. Broadband IDT using Fourier synthesized design can also be used to generate sharp SAW pulses in a scheme which recently has been proposed as an ultrafast way to modulate quantum dots in photonic nanocavities for quantum photonics[37].

**Discussion**

In conclusion, acousto-optic modulation of photonic resonators with surface acoustic wave at frequencies above 10 GHz has been demonstrated. The acousto-optic system is completely integrated on piezoelectric AlN film deposited on silicon based substrate so it is potentially compatible with the silicon photonics platform. With the achieved ultrahigh acoustic frequency, an unprecedented sub-optical wavelength regime of acousto-optics is reached. In this regime, the efficiency of light-sound interaction depends strongly on the modal and phase matching between the acoustic and optical modes, and can be optimized with proper device design to achieve optimal modulation efficiency and ultrahigh modulation frequency. On this highly integrated acousto-optic platform, we expect that a wide range of novel device applications may be developed, including ultrafast optical modulators, non-reciprocal photonic devices based on Brillouin scattering[38-40] and side-band resolved cavity optomechanics utilizing actively excited travelling acoustic waves[41-43].

## Method

## Device Fabrication

The integrated surface acousto-optic devices were fabricated on 330 nm thick c-axis oriented polycrystalline piezoelectric AlN thin film sputtered (by OEM Group, AZ) on silicon wafers with a 3µm thermally grown oxide layer. The photonics layer was patterned with electron beam lithography (Vistec EBPG-5000+) using ZEP-520 resist and etched with $Cl_2$ based reactive ion etching. The AlN rib waveguides were etched 200nm deep with 130nm thick slabs. The slab allows the acoustic wave to propagate across the waveguide with reduced reflection and transmission loss. The IDTs to excite SAW were patterned with electron beam lithography using ZEP-520 resist followed by deposition of 40nm thick chrome/gold film and a liftoff process.

## Measurement Methods

The racetrack resonator was characterized by measuring the transmission using a tunable diode laser (Agilent 81940A). The SAW IDTs were characterized with a PNA vector network analyzer (Agilent E8362B) via a microwave probe and measuring the corresponding $S_{11}$ reflection spectrum. Frequency response in both magnitude and phase were recorded in order to determine the electromechanical coupling efficiency of the devices. This measurement was performed on a spectrum of SAW devices of varying wavelengths to determine the dispersion relation of the devices. Before every measurement, a calibration substrate (GGB Inc.) was used to calibrate and null off the impedance of the cables and the microwave probes. Acousto-optic modulation was measured with the slope detection scheme with the laser wavelength detuned from the optical resonance. The optical signal from the device was first amplified by an erbium doped fiber amplifier (EDFA) and filtered with a tunable optical filter to remove amplified spontaneous emission noise induced by the EDFA. Finally the signal was sent to a high speed photoreceiver (12 GHz bandwidth, New Focus 1554A). The photoreciever output was amplified using a low noise amplifier and then connected to the vector network analyzer to measure the optical $S_{21}$ frequency response when the network analyzer source frequency was swept. The measurement schematics is shown in Supplementary Figure 5.

**Author Contributions**

M. L. conceived and supervised the research; S. A. T. and M. L. designed the experiments; S. A. T. performed the fabrication and measurement; S. A. T. analyzed the data; M. L. and S. A. T. co-wrote the paper.

**Acknowledgement:**

We acknowledge the funding support provided by the National Science Foundation (Award No. ECCS-1307601) and the Young Investigator Program (YIP) of AFOSR (Award No. FA9550-12-1-0338). Parts of this work were carried out in the University of Minnesota Nanofabrication Center which receives partial support from NSF through NNIN program, and the Characterization Facility which is a member of the NSF-funded Materials Research Facilities Network via the MRSEC program.

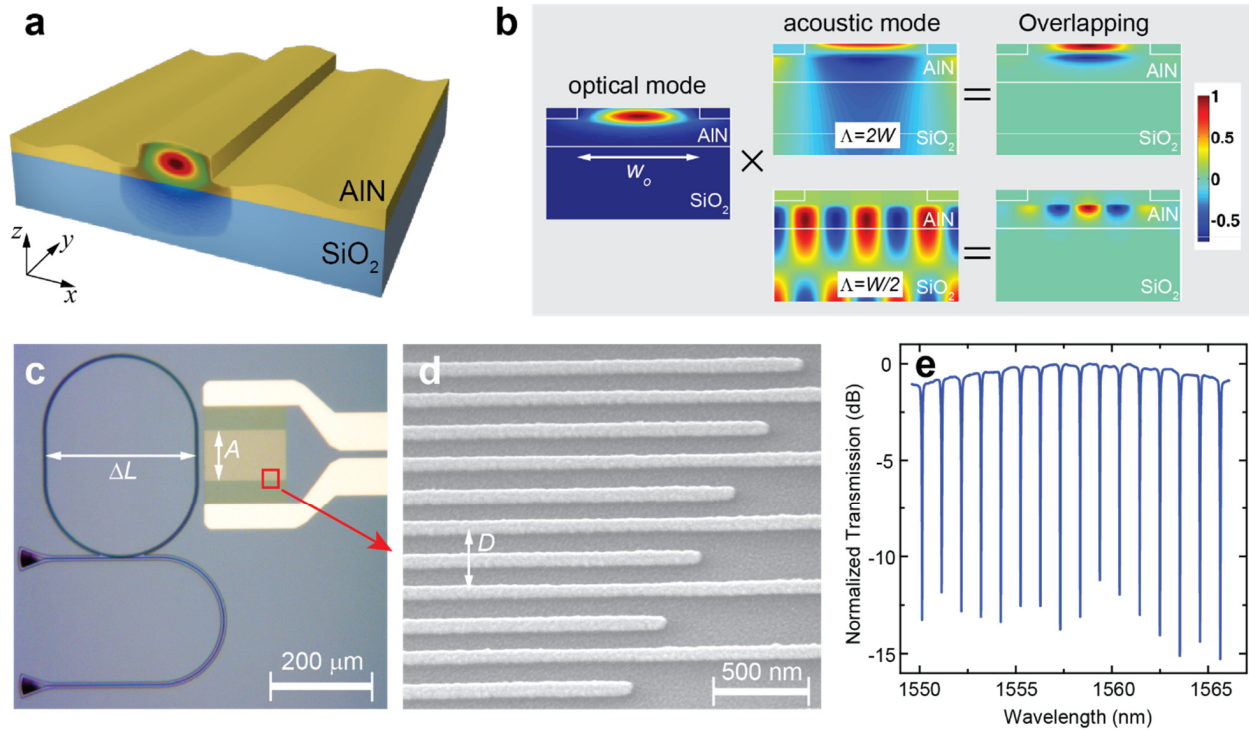

**Figure 1 An integrated surface acousto-optic system. a.** Schematics illustrating the relative scale and the interaction between the surface acoustic wave and the optical mode of the integrated waveguide made of AlN on a SiO$_2$ cladding layer. Overlaid on the facet of the waveguide is simulated optical mode. **b.** Numerical simulation results of the field distribution of the optical mode, the surface acoustic mode and their overlapping as defined in the text for two representative situations when $\Lambda=2W$ and $\Lambda=W/2$ . **c.** Optical microscope image of a typical device, featuring a racetrack resonator made of AlN rib waveguide and an inter-digital transducer (IDT). The aperture size of the IDT is $A$ and the distance between the front and back straight segments of the racetrack is $\Delta L$ as shown by the labels. **d.** Scanning electron microscope image of the electrode fingers of an IDT with a period $D$ of 400 nm. The width of the fingers is 100 nm. **e.** Transmission spectrum of the racetrack resonator measured from the feeding waveguide with integrated grating couplers. The waveguide loaded optical quality factor is $4\times10^4$, corresponding to an intrinsic value of $8\times10^4$.

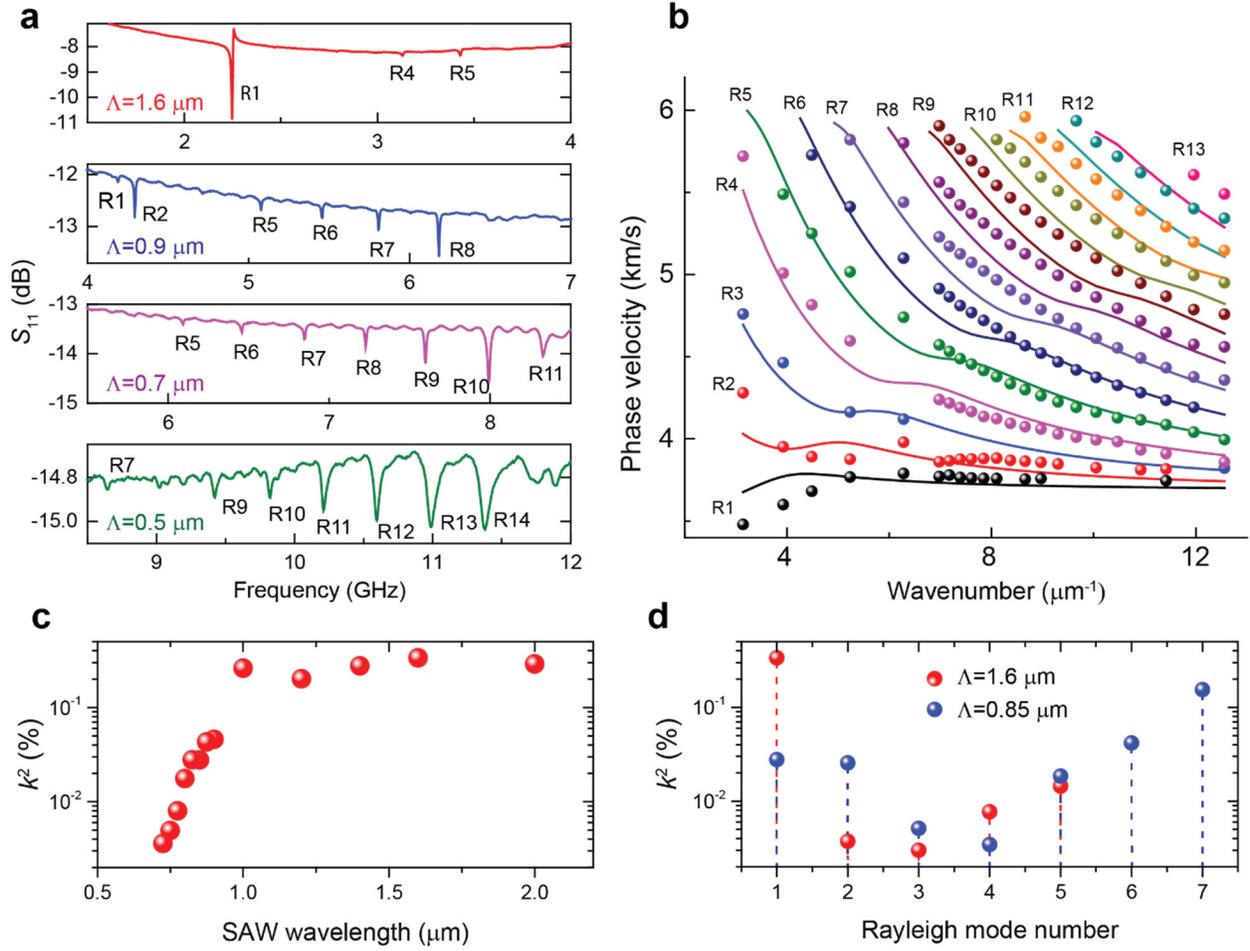

**Figure 2 Surface acoustic wave characteristics. a.** Measured spectra of $S_{11}$ reflection coefficient of IDTs with different wavelengths. The excited acoustic modes appear as dips in the spectra, which are labeled with the corresponding Rayleigh mode orders. **b.** Dispersion relation between the phase velocities of different Rayleigh modes and the wavenumber. Lines are simulated results and symbols are experimentally determined values. **c.** Electromechanical coupling coefficient $k^2$ determined from the measured $S_{11}$ reflection spectrum of the R1 modes of IDTs with various wavelengths. **d.** Electromechanical coupling coefficients of different modes of two IDT devices with $\Lambda$=1.6μm (red symbols) and 0.85μm (blue symbols).

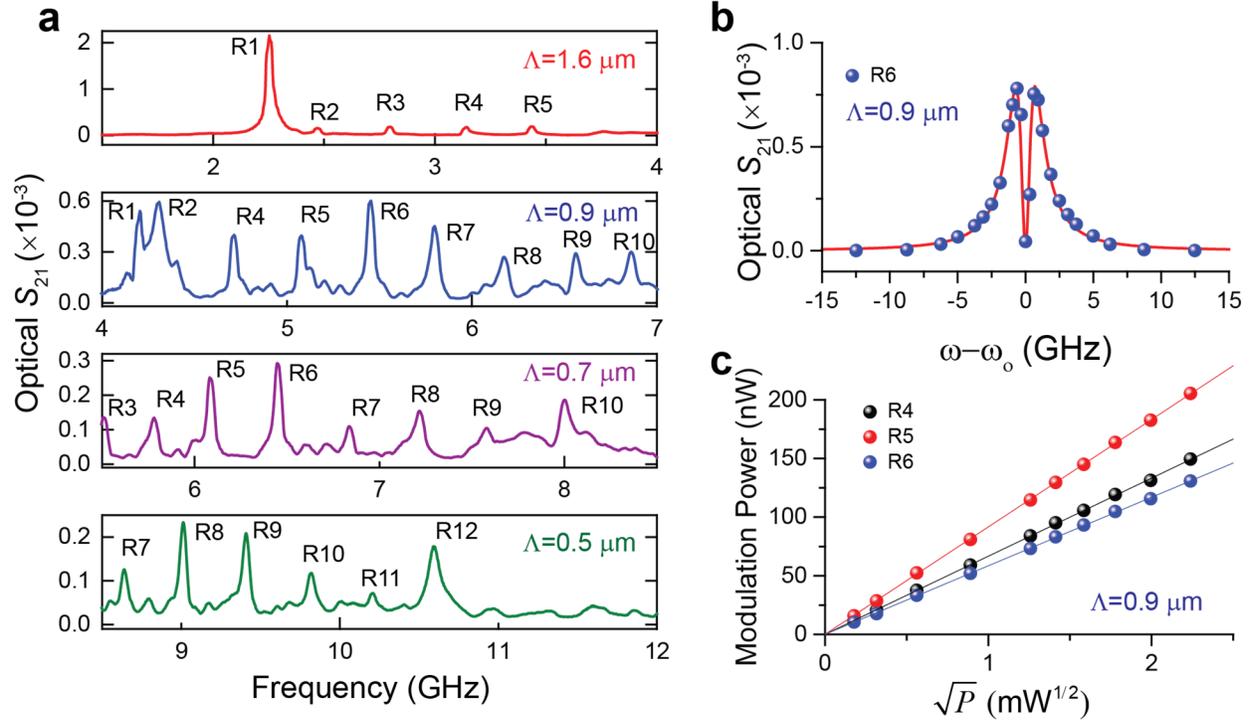

**Figure 3 Ultrahigh frequency acoustic modulation of optical resonators. a.** Measured spectra of $S_{21}$ transmission coefficient of the integrated acousto-optic devices that are the same as in Fig. 2a. Modulation of the photonic resonances shows as peaks in the spectra which are labeled with the corresponding Rayleigh mode orders. The width of the waveguide is constantly 0.8 μm in these devices. **b.** Acousto-optic modulation response versus laser detuning relative to the optical resonance for the R6 mode in the $\Lambda$=0.9 μm device. The response shows a line shape of the derivation of a Lorentzian resonance, indicating that the acoustic wave modulates the phase of the optical mode. **c.** Modulated optical signal power versus the square root of the input power and linear fittings for the R4, R5 and R6 modes of the $\Lambda$=0.9 μm device. The linear dependence indicates that the modulation is proportional to the amplitude of the acoustic wave.

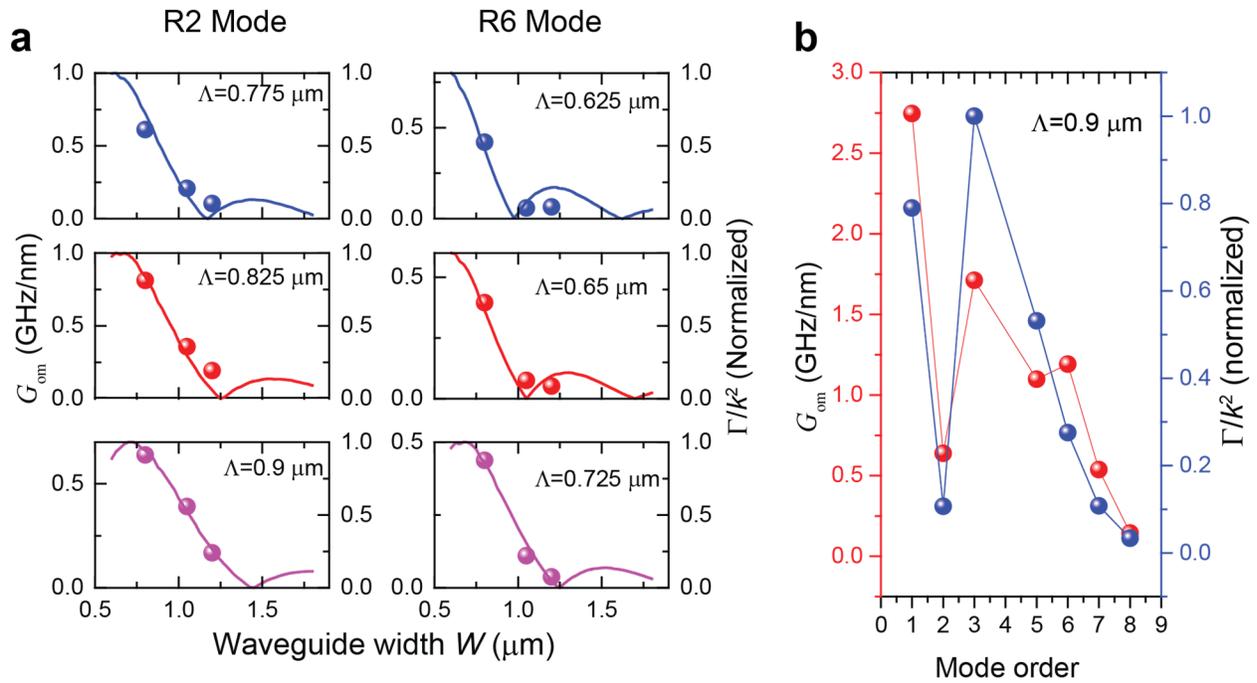

**Figure 4 Optomechanical coupling rate and modal overlap between acoustic and optical modes. a.** Measured optomechanical coupling coefficient $G_{om}$ and theoretically calculated overlap factor $\Gamma$ of the acoustic and the optical modes after normalization by the electromechanical coupling coefficient $k^2$, for the R2 and R6 modes of devices with varying acoustic wavelength $\Lambda$. Note the experimental (symbols) and the theoretical results (lines) use different scales but qualitatively they agree very well. **b.** Measured $G_{om}$ (red symbols) and theoretically calculated $\Gamma/k^2$ (blue symbols) of all the acoustic mode for the device with $\Lambda$=0.9 μm.

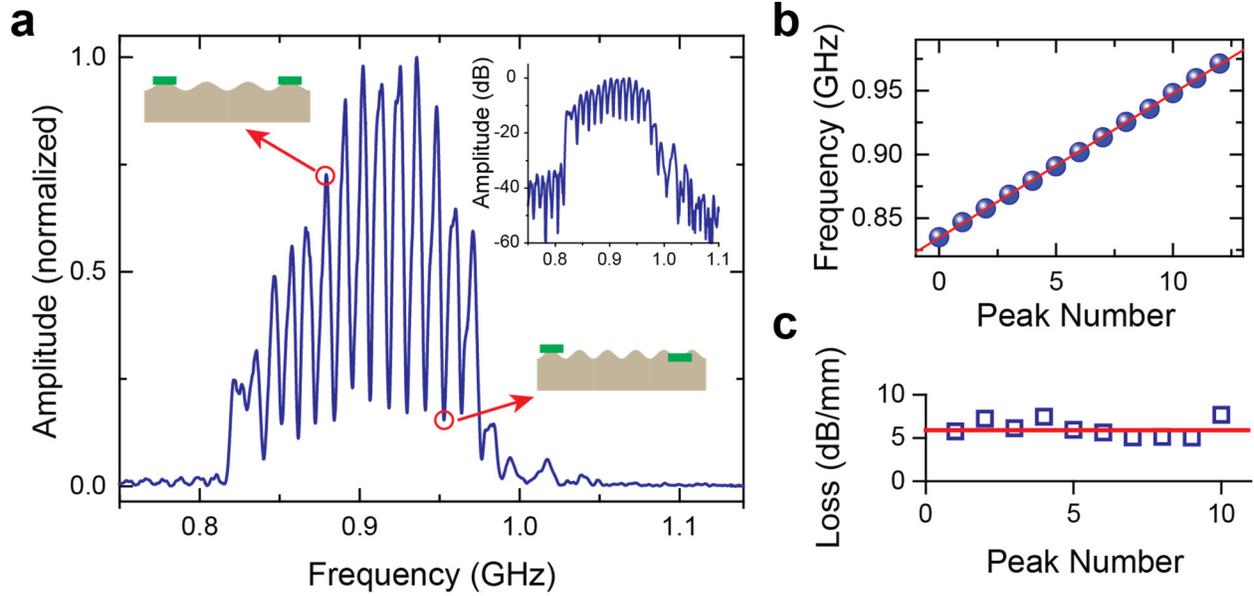

**Figure 5 Broadband acousto-optic modulation. a.** Frequency response of the optical signal to broadband acoustic modulation generated by a slanted IDT with a center period of 4 µm. The −3 dB bandwidth of the modulation is 130 MHz and the stopband rejection ratio is higher than 30 dB (upper-right inset). Strong ripples seen in the passband can be attributed to the delay of the acoustic wave at the front and back waveguide segments of the racetrack resonator (upper-left and lower-right insets) the interference of the modulation. **b.** The frequencies of the peaks in the passband ripples increase linearly with the peak number. Linear fitting yields the phase velocity of 3400 m/s for the fundamental Rayleigh mode, in agreement with previous results. **c.** From the extinction ratio of the ripple, the propagation loss of the acoustic wave can be estimated. The result suggests a loss of 6 dB/mm, which should be considered as an upper bound of the loss.